\documentclass{aa}  

\usepackage{graphicx}
\usepackage{txfonts}
\usepackage{hyperref}
\begin{document}

   \title{A multilevel implementation of the Goldreich-Kylafis effect into the radiative transfer code PyRaTE}

   \author{A. Tritsis
          \inst{1}
          \and
          N. Kylafis\inst{2, 3}
          }

   \institute{Institute of Physics, Laboratory of Astrophysics, Ecole Polytechnique F\'ed\'erale de Lausanne (EPFL), \\ Observatoire de Sauverny, 1290, Versoix, Switzerland \\
              \email{aris.tritsis@epfl.ch}
         \and
             University of Crete, Physics Department \& Institute of Theoretical \& Computational Physics, 70013 Heraklion, Greece
        \and
             Institute of Astrophysics, Foundation for Research and Technology -- Hellas, 70013 Heraklion, Greece}

   \date{Received date; accepted date}
   \titlerunning{Mock observations of the GK effect}
   \authorrunning{Tritsis \& Kylafis}
 
  \abstract
   {Among all the available observational techniques for studying magnetic fields in the dense cold phase of the interstellar medium, linear polarization of spectral lines, referred to in the literature as the Goldreich-Kylafis effect (Goldreich \& Kylafis 1981; hereafter ``GK effect''), remains one of the most underutilized methods.}
   {In this study, we implement the GK effect into the multilevel, non-local thermodynamic equilibrium radiative transfer code \textsc{PyRaTE}.}
   {Different modes of polarized radiation are treated individually with separate optical depths computed for each polarization direction. We benchmark our implementation against analytical results and provide tests for various limiting cases.}
   {In agreement with previous theoretical results, we find that in the multilevel case the amount of fractional polarization decreases when compared to the two-level approximation, but this result is subject to the relative importance between radiative and collisional processes. Finally, we post-process an axially symmetric, non-ideal magnetohydrodynamic chemo-dynamical simulation of a collapsing prestellar core and provide theoretical predictions regarding the shape (as a function of velocity) of the polarization fraction of $\rm{CO}$ during the early stages in the evolution of molecular clouds. The code is freely available to \href{https://github.com/ArisTr/PyRaTE.git}{download}.}
   {}


   \maketitle


\section{Introduction}\label{intro}

The magnetic field is one of the key ingredients required in our efforts to understand the dynamical evolution of molecular clouds and prestellar cores (Mouschovias \& Ciolek 1999). Observationally, our knowledge of the magnetic field in molecular clouds mostly stems from polarized thermal dust emission (e.g., Ward-Thompson et al. 2017) and dust-induced starlight polarization (e.g., Panopoulou et al. 2016). Both of these techniques probe the orientation of the plane-of-sky (POS) component of the magnetic field, while its strength can be measured under certain assumptions regarding the equipartition between the kinetic and magnetic energy (Davis 1951; Chandrasekhar \& Fermi 1953; see also Skalidis \& Tassis 2021 and references therein).

On the other hand, the line-of-sight component of the magnetic field can be probed by Zeeman circular polarization measurements of molecular spectral lines (e.g. Troland \& Crutcher 2008; Falgarone et al. 2008). The Zeeman effect remains the only available method that can directly yield the strength of the magnetic field (as well as its direction) and it is therefore extremely valuable. However, due to the very high signal-to-noise ratio required, robust measurements of the Zeeman effect can be extremely challenging, often requiring as much as $\sim$10 hours of integration time per pointing (Crutcher et al. 2009).

On the subject of radio observations of molecular spectra, one of the most under-explored techniques for probing the direction of magnetic fields is the so-called Goldreich-Kylafis effect (hereafter ``GK effect''; Goldreich \& Kylafis 1981). The GK effect refers to linear polarization of molecular spectral lines and arises when the magnetic sublevels are unequally populated due to an anisotropic velocity field. For the GK effect to arise, the optical depth of the line needs to be moderate and anisotropic and the radiative rates need to be comparable to the collisional ones for excitation and de-excitation. Therefore, the GK effect arises only under non-local thermodynamic equilibrium (non-LTE) conditions. Linear polarization of spectral lines is also shown to arise in masers (Goldreich et al. 1973; Deguchi \& Watson 1990; Lankhaar \& Vlemmings 2019), in circumstellar envelopes through directional continuum emission in evolved stars (Morris et al. 1985), and in star-forming regions via directional collisions (Lankhaar \& Vlemmings 2020). Finally, linearly polarized radiation can be transformed to circular polarized through the so-called Anisotropic Resonant Scattering effect (Houde et al. 2022 and references therein).

In the prestellar phase of molecular clouds, only a handful of observational surveys of the GK effect have been reported to date (Lai et al. 2003; Girart et al. 2004; Cortes et al. 2005; Forbrich et al. 2008; Cort{\'e}s et al. 2021; Barnes et al. 2023). However, on the theoretical front, the full theoretical formalism for modelling the GK effect has been developed more than four decades ago (Goldreich \& Kylafis 1981; Goldreich \& Kylafis 1982; Kylafis 1983; Deguchi \& Watson 1984; Cortes et al. 2005; Yang \& Lai 2010; Huang et al. 2020). Lankhaar \& Vlemmings (2020) recently developed the \textsc{PORTAL} (POlarized Radiative Transfer Adapted to Line) radiative-transfer code, which builds upon the non-polarized results from the \textsc{LIME} (Line Modeling Engine; Brinch \& Hogerheijde 2010) radiative-transfer code to yield the polarization fraction. However, in \textsc{PORTAL}, only the anisotropy of the total radiation is considered, instead of the two polarization modes of the radiation being considered individually. Even though Lankhaar \& Vlemmings (2020) demonstrated the validity of their approximation in specialized cases where they found deviations from the analytical results of Kylafis (1983) only for high optical depths, it remains ambiguous how well their approximation performs in the general case.

Here, we modify the \textsc{PyRaTE} (Python Radiative Transfer Emission) code (Tritsis et al. 2018) to include the GK effect following the theoretical formalism by Deguchi \& Watson (1984). Under this multilevel formalism, the different modes of polarized radiation are considered individually. In contrast to other radiative-transfer codes (e.g. \textsc{RADEX}; van der Tak et al. 2007) the optical depth in \textsc{PyRaTE} is computed more accurately by taking into account variations in all relevant physical quantities i.e. the $\rm{H_2}$ number density, the molecular number density, the temperature and the velocity structure of the physical system under consideration.

This study is organized as follows: In section \S~\ref{numer} we outline the theoretical background and basic equations. In \S~\ref{bench} we benchmark our code by performing various tests for limiting cases and by comparing our numerical calculations against analytical results. In \S~\ref{GKrealistic} we present polarized line radiative-transfer calculations from a non-ideal magnetohydrodynamic (MHD) simulation of a collapsing prestellar core, where the number density of $\rm{CO}$ is computed on-the-fly using a non-equilibrium chemical model. Finally, we summarize our results and conclude in \S~\ref{discuss}.


\section{Basic equations and numerical implementation}\label{numer}

\begin{figure*}
\includegraphics[width=2.1\columnwidth, clip]{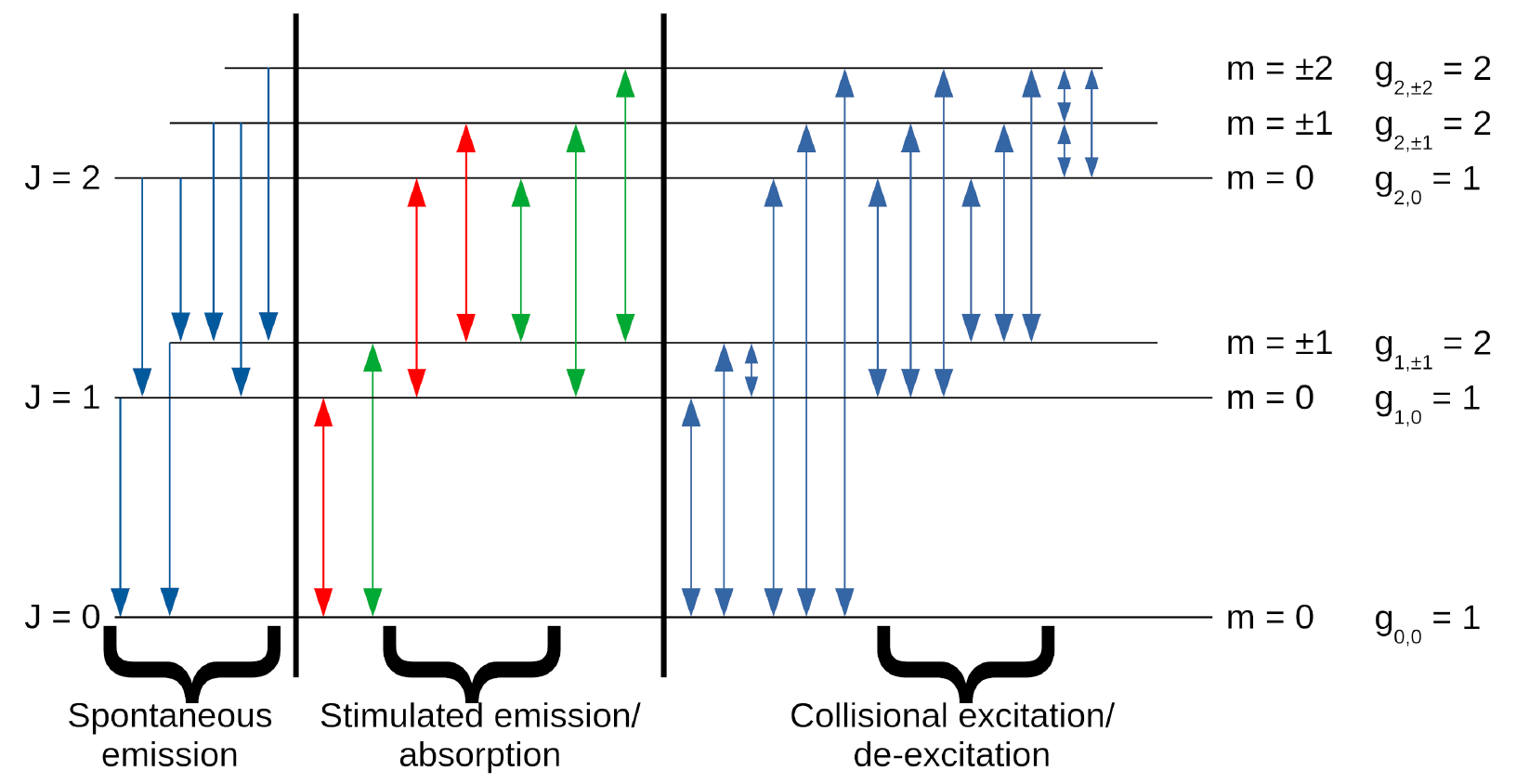}
\caption{Schematic representation of all the processes considered for a molecular species with three rotational energy levels, when the magnetic sublevels are considered independently. Absorption and stimulated emission associated with $\rm{\Delta m}=0$ are marked with red, while the corresponding transitions associated with $\rm{\lvert \Delta m \rvert}=1$ are marked with green color.
\label{schematic}}
\end{figure*}

Figure~\ref{schematic} schematically depicts all the processes that need to be considered for a linear molecule with three rotational levels, when the magnetic sublevels are considered individually. Here, we have separated between spontaneous emission, absorption and stimulated emission, and collisional excitation and de-excitation processes. Compared to the case where the population densities are (2$J$+1) times degenerate, the level of complication increases substantially. Specifically, we need to consider all the spontaneous and stimulated emission/absorption processes allowed by the selection rules ($\rm{\Delta \textit{J}}=1$, $\rm{\lvert \Delta m \rvert}=0, 1$) separately. Additionally, for collisional processes we need to consider all the possible transitions between every sublevel. Most significantly however, for each pair of rotational levels $J$ and $J-1$ (henceforth denoted as $J'$), two sets of escape probabilities need to be computed (one for each polarization mode), instead of one. To make matters even more complicated, the escape probability associated with the specific intensity polarized parallel to the magnetic field ($I^{\parallel}_{J, J'}$) depends upon the angle between the direction of propagation of radiation with the magnetic field (henceforth denoted as $\gamma$). 

In the following, the net radiative rates associated with transitions with $\rm{\Delta m}=0$ and $\rm{\lvert \Delta m \rvert}=1$ are denoted respectively with $R_{J,J'}$ and $U_{J,J'}$. For the rest of the quantities, we follow the notation by Deguchi \& Watson (1984). The quantities $R_{J,J'}$ and $U_{J,J'}$ are given by
\begin{subequations}\label{eq:radrates}
\begin{equation}\label{EqRjj}
R_{J,J'} = 3 B_{\textit{J,m}\rightarrow \textit{J}',\textit{m}'} \int \frac{d\Omega}{4\pi} \sin^2\gamma \int d\nu \phi (\nu - \nu_{J, J'}) I^{\parallel}_{J, J'}(\vec{\hat{\Omega}}),
\end{equation}

\begin{eqnarray}\label{EqUjj}
U_{J,J'} = \frac{3B_{\textit{J,m}\rightarrow \textit{J}',\textit{m}'}}{2}\int \frac{d\Omega}{4\pi} \int d\nu \phi (\nu - \nu_{J, J'})  \nonumber \\ 
 \{I^{\perp}_{J, J'}(\vec{\hat{\Omega}}) + \cos^2\gamma I^{\parallel}_{J, J'}(\vec{\hat{\Omega}})\},
\end{eqnarray}
\end{subequations}
where $\nu$ and $\nu_{J, J'}$ are, respectively, the frequency, and the rest frequency of the line. Furthermore, $\vec{\hat{\Omega}}$ is the unit vector of the solid angle, $B_{\textit{J,m}\rightarrow \textit{J}',\textit{m}'}$ is the Einstein B coefficient for stimulated emission, $\phi (\nu - \nu_{J, J'})$ is the normalized profile function and $I^{\perp}_{J, J'}$ denotes the specific intensity of radiation polarized perpendicular to the magnetic field.

Under the large-velocity-gradient (LVG) approximation (Sobolev 1960; Castor 1970; Lucy 1971), we have that
\begin{equation}\label{lvgapprox}
\int d\nu \phi (\nu - \nu_{J, J'}) I^{q}_{J, J'} = S^{q}_{J, J'}(1-\beta^{q}_{J, J'}) + \frac{B}{2}\beta^{q}_{J, J'},
\end{equation}
where $q = \parallel$ or $\perp$, $\beta$ is the probability that a photon escapes the cloud, and $S^{q}_{J, J'}$ is the source function. The second term on the right-hand side of Eq.~(\ref{lvgapprox}) represents the contribution due to external photons penetrating the cloud. Here, we assume that the only external radiation is due to the cosmic microwave background (CMB; henceforth denoted as $B_{CMB}$). Following Cortes et al. (2005), a compact external continuum source can also be added to the code by setting $B = B_{CMB} + S(\vec{\hat{\Omega}})$ where
\begin{equation}\label{cntSource}
S(\vec{\hat{\Omega}})=(1-e^{-\tau_c})B_\nu(T_{source}).
\end{equation}
In Eq.~\ref{cntSource}, $B_\nu(T_{source})$ and $T_{source}$ are, respectively, Planck's function and the temperature of the continuum source, and $\tau_c$ is the optical depth for continuum emission from the source. For further details on adding an external continuum source we refer the reader to Cortes et al. (2005). The escape probability is related to the optical depth of the line as $\beta^{q}_{J, J'} = (1+e^{-\tau^q_{J', J}})/\tau^q_{J', J}$ (Mihalas 1978; de Jong et al. 1980\footnote{For a comparison of the escape probability computed as a function of the optical depth under different assumptions for the geometry of the cloud, we refer the reader to van der Tak et al. (2007).}). Although, in strict terms, the latter equation holds true only under the LVG approximation, one can still physically expect that an escape probability can be defined even when the velocity gradients in the physical system of interest are not large enough for the LVG to be valid. In turn, the optical depth of the radiation polarized parallel and perpendicular to the magnetic field can be computed by integrating the absorption coefficient $\kappa^q_{J',J}$ which is given by
\begin{subequations}\label{eq:absorpParPerp}
\begin{equation}\label{absorpPepr}
\kappa^\perp_{J',J} = \frac{1}{2}\phi (\nu - \nu_{J, J'}) \sum_{\Delta m =1} \kappa_{\textit{J}',\textit{m}'\rightarrow \textit{J,m}}
\end{equation}

\begin{eqnarray}\label{absorpParal}
\kappa^\parallel_{J',J} = \phi (\nu - \nu_{J, J'}) (\sin^2\gamma\sum_{\Delta m =0} \kappa_{\textit{J}',\textit{m}'\rightarrow \textit{J,m}} \nonumber \\
\ +\frac{1}{2} \cos^2\gamma\sum_{\Delta m =1} \kappa_{\textit{J}',\textit{m}'\rightarrow \textit{J,m}}).
\end{eqnarray}
\end{subequations}
In Eqs.~(\ref{absorpPepr}) and~(\ref{absorpParal}), $\kappa_{\textit{J}',\textit{m}'\rightarrow \textit{J,m}}$ is computed as
\begin{eqnarray}\label{absorp}
\kappa_{\textit{J}',\textit{m}'\rightarrow \textit{J,m}} = \frac{3}{8\pi}\Big(\frac{c}{\nu_{J, J'}}\Big)^2 A_{\textit{J,m}\rightarrow \textit{J}',\textit{m}'} \max(g_{\textit{J,m}}, g_{\textit{J}',\textit{m}'}) \nonumber \\
\ (n_{\textit{J}',\textit{m}'} - n_{\textit{J,m}}),
\end{eqnarray}
where $c$ is the speed of light, $A_{\textit{J,m}\rightarrow \textit{J}',\textit{m}'}$ is the Einstein $A$ coefficient and $n_{\textit{J}',\textit{m}'}$ and $n_{\textit{J,m}}$ are the level populations of the lower and upper energy sublevels, respectively. The level populations can be computed by solving the detailed balance equations (see Appendix~\ref{StEqEq}). While solving the detailed balance equations and throughout the code, we impose that $n_{J, m} = n_{J, -m}$ (for $m\neq 0$). The term $\max(g_{\textit{J,m}}, g_{\textit{J}',\textit{m}'})$ that appears in Eq.~(\ref{absorp}) ensures, however, that every process that needs to be considered twice (e.g. ($\textit{2}, \lvert\textit{2}\rvert) \rightarrow (\textit{1},\lvert\textit{1}\rvert$) or in other words, from level ($\textit{2}, \textit{2}$) to ($\textit{1},\textit{1}$) and from level ($\textit{2}, \textit{-2}$) to level ($\textit{1},\textit{-1}$), is correctly being done so. Finally, the source functions $S^{q}_{J, J'}$ that appears in Eq.~(\ref{lvgapprox}) are given by
\begin{subequations}\label{eq:sourceParPerp}
\begin{equation}\label{sourcePepr}
S^\perp_{J,J'} = \frac{\sum_{\Delta m = 1} \kappa_{\textit{J}',\textit{m}'\rightarrow \textit{J,m}} S_{\textit{J,m}\rightarrow \textit{J}',\textit{m}'}}{\sum_{\Delta m = 1} \kappa_{\textit{J}',\textit{m}'\rightarrow \textit{J,m}}},
\end{equation}

\begin{eqnarray}\label{sourceParal}
S^\parallel_{J,J'} = \Big(\sin^2\gamma\sum_{\Delta m = 0} \kappa_{\textit{J}',\textit{m}'\rightarrow \textit{J,m}} S_{\textit{J,m}\rightarrow \textit{J}',\textit{m}'} \nonumber \\ 
\ + \frac{1}{2}\cos^2\gamma\sum_{\Delta m = 1} \kappa_{\textit{J}',\textit{m}'\rightarrow \textit{J,m}} S_{\textit{J,m}\rightarrow \textit{J}',\textit{m}'}\Big) \nonumber \\  
\ \Big(\sin^2\gamma\sum_{\Delta m = 0} \kappa_{\textit{J}',\textit{m}'\rightarrow \textit{J,m}} + \frac{1}{2}\cos^2\gamma\sum_{\Delta m = 1} \kappa_{\textit{J}',\textit{m}'\rightarrow \textit{J,m}} \Big)^{-1},
\end{eqnarray}
\end{subequations}
where
\begin{eqnarray}\label{sourcef}
S_{\textit{J,m}\rightarrow \textit{J}',\textit{m}'} = \frac{h\nu_{J, J'}^3}{c^2} \frac{n_{\textit{J,m}}}{n_{\textit{J}',\textit{m}'} - n_{\textit{J,m}}}.
\end{eqnarray}
In Eq.~\ref{sourcef}, $h$ is Planck's constant.

\begin{figure}
\includegraphics[width=1\columnwidth, clip]{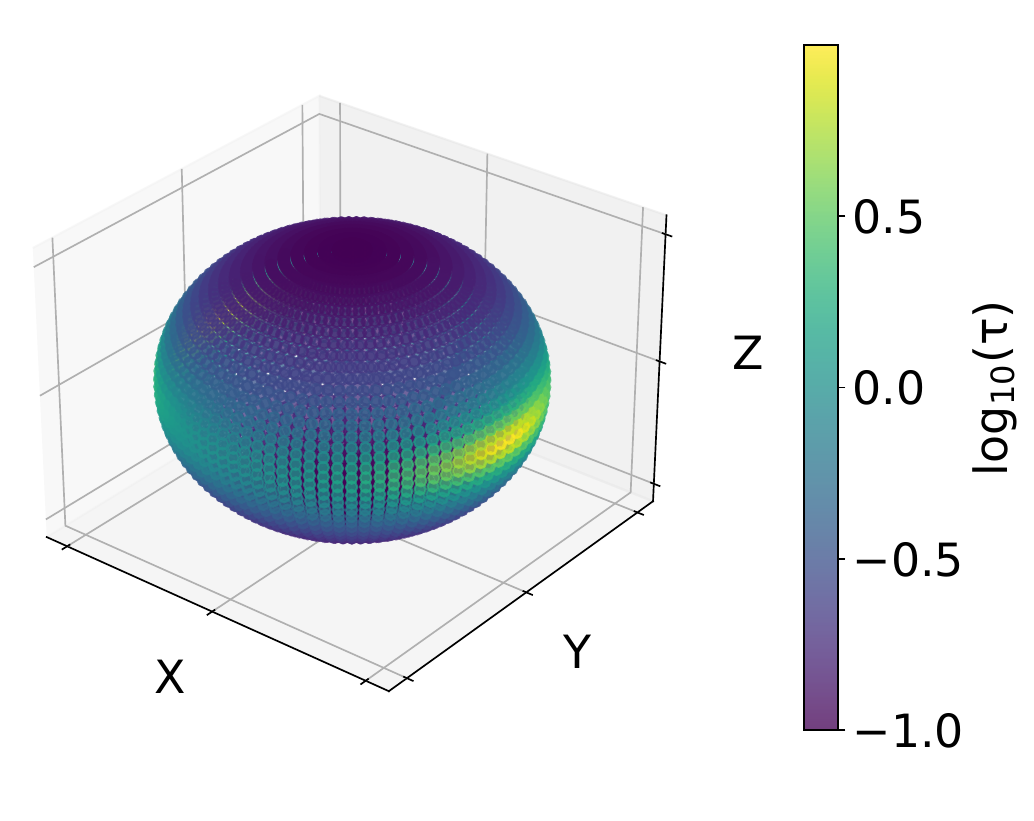}
\caption{Interpolated values of the optical depth inside a single cell, based on Eq.~(\ref{InterpBeta}), along each direction $\vec{\hat{\Omega}}$. The values of the optical depth along the principle axes are set to $\tau^{x^+} = \tau^{x^-} = 10$, $\tau^{y^+} = \tau^{y^-} = 1$ and $\tau^{z^+} = \tau^{z^-} = 0.1$.
\label{InterpBetasFig}}
\end{figure}

The optical depth $\tau^q_{J', J}$ for each polarization direction for a grid point ($i', j', k'$),  is calculated along both directions of the principal axes of our numerical grid, by adding the absorption coefficient of the grid points for which their velocity difference with grid point ($i', j', k'$) is less than the thermal linewidth (see also the discussion in Appendix~\ref{SupersedingLVG}). For instance, the optical depth along the $x^+$ and $x^-$ directions is
\begin{subequations}\label{eq:tauQpQm}
\begin{eqnarray}\label{tauQp}
\tau^{q, x^+}_{J', J} = \Big(\frac{\kappa_{J', J, i'j'k'}^q}{2}+ \nonumber \\
\sum\limits_{i=i^\prime+1}^{X} \kappa_{J', J, ijk}^q[\mid \varv_{i^\prime j^\prime k^\prime}-\varv_{ij^\prime k^\prime}\mid<\Delta \varv_{ij^\prime k^\prime}^{th}]\Big) \Delta x
\end{eqnarray}
\begin{eqnarray}\label{tauQm}
\tau^{q, x^-}_{J', J} = \Big(\frac{\kappa_{J', J, i'j'k'}^q}{2}+ \nonumber \\
\sum\limits_{i=i^\prime-1}^{0} \kappa_{J', J, ijk}^q[\mid \varv_{i^\prime j^\prime k^\prime}-\varv_{ij^\prime k^\prime}\mid<\Delta \varv_{ij^\prime k^\prime}^{th}]\Big) \Delta x,
\end{eqnarray}
\end{subequations}
where $X$ is the size of our grid in the $x$ direction, $\Delta x$ is the size of the cell in the same direction, $\varv$ is the velocity and $\Delta \varv^{th}$ is the thermal linewidth. An identical process is adopted for the $y$ and $z$ directions. Consequently, for each pair of rotational levels $J, J'$ we compute six optical depths for each polarization direction. Then, we numerically solve the integrals in Eqs.~(\ref{EqRjj} \&~\ref{EqUjj})\footnote{Integration is performed using the Quadpack library (Piessens et al. 1983).} by computing a value for the optical depth for every direction $\vec{\hat{\Omega}}$ as
\begin{equation}\label{InterpBeta}
\frac{1}{\tau(\vec{\hat{\Omega}})} = \frac{\sum\limits_{n=1}^6 \frac{1}{\tau_n} (\vec{\hat{\Omega}}\cdot{\vec{\hat{w}_n}}[\vec{\hat{\Omega}}\cdot{\vec{\hat{w}_n}} > 0])^2}{\sum\limits_{n=1}^6 (\vec{\hat{\Omega}}\cdot\vec{{\hat{w}_n}}[\vec{\hat{\Omega}}\cdot\vec{{\hat{w}_n}} > 0])^2},
\end{equation}
where $\vec{\hat{w}_n}$ are the unit vectors along the principle axes (both directions) and $\tau_n$ are the values of the optical depth along these directions. As an example, we show in Fig.~\ref{InterpBetasFig} the interpolated values for the optical depth when the values of the optical depth along the principle axes are $\tau^{x^+} = \tau^{x^-} = 10$, $\tau^{y^+} = \tau^{y^-} = 1$ and $\tau^{z^+} = \tau^{z^-} = 0.1$.

For the Einstein $A$ coefficient that first appeared in Eq.~(\ref{absorp}), Deguchi \& Watson (1984) reference Townes \& Schawlow (1955) and adopt
\begin{equation}\label{EinAwrong}
A_{\textit{J,m}\rightarrow \textit{J}',\textit{m}'} = \begin{cases} \frac{(J+1)^2 - m^2}{(2J+1)(J+1)}~A_{\textit{J}\rightarrow \textit{J}'}, & \text{for $\Delta m = 0$} \\
\\
\frac{(J+\lvert m \rvert+1)(J+\lvert m \rvert)}{2(2J+1)(J+1)}~A_{\textit{J}\rightarrow \textit{J}'}, & \text{for $\lvert \Delta m \rvert = 1$},
\end{cases}
\end{equation}
where the value of $A_{\textit{J}\rightarrow \textit{J}'}$ can be found in databases for molecular spectroscopy such as the $\textsc{LAMBDA}$ database (Sch{\"o}ier et al. 2005). However, the expression given in Eq.~(\ref{EinAwrong}) is $\textit{not}$ correct and should not be used as it leads to non-negligible linear polarization (of the order of a few \%) even under LTE conditions. Instead, the Einstein $A$ coefficient should be computed as
\begin{equation}\label{EinAcorrect}
A_{\textit{J,m}\rightarrow \textit{J}',\textit{m}'} = \begin{pmatrix}
\textit{J} & 1 & \textit{J}'\\
\textit{-m} & \textit{m} - \textit{m}' & \textit{m}'
\end{pmatrix}^2 A_{\textit{J}\rightarrow \textit{J}'},
\end{equation}
where the term in the brackets is the Wigner 3$j$ symbol (Stenflo 1994)\footnote{The Wigner 3$j$ symbol is calculated using the \textsc{sympy} \textsc{python} package (Rasch \& Yu  2003).}. For the transition $J = 1\rightarrow 0$, Eq.~(\ref{EinAcorrect}) gives that $A_{\textit{1,0}\rightarrow \textit{0},\textit{0}}$ = $A_{\textit{1,}\pm \textit{1}\rightarrow \textit{0},\textit{0}}$ = $\frac{1}{3} A_{\textit{1}\rightarrow \textit{0}}$, meaning that spontaneous transitions between all three sublevels are equally likely\footnote{The same however is not true for the transition $J = 2\rightarrow 1$ where, for instance, the transition ($\textit{2}, \lvert\textit{2}\rvert) \rightarrow (\textit{1},\lvert\textit{1}\rvert$) is twice as likely than the transition ($\textit{2}, \lvert\textit{1}\rvert) \rightarrow (\textit{1},\lvert\textit{1}\rvert$).}. On the other hand, Eq.~(\ref{EinAwrong}) gives $A_{\textit{1,0}\rightarrow \textit{0},\textit{0}}$ = $\frac{2}{3}A_{\textit{1}\rightarrow \textit{0}}$ and $A_{\textit{1,}\pm \textit{1}\rightarrow \textit{0},\textit{0}}$ = $\frac{1}{2}A_{\textit{1}\rightarrow \textit{0}}$. Clearly, adopting Eq.~(\ref{EinAwrong}) instead of Eq.~(\ref{EinAcorrect}) will lead to the magnetic sublevels of $J = 1$ having unequal populations, even without stimulated processes taken into account.

For the collisional coefficients we follow Deguchi \& Watson (1984) and adopt
\begin{equation}\label{collcoeffs}
C_{\textit{J,m}\rightarrow \textit{J}',\textit{m}'} = \frac{C_{\textit{J}\rightarrow \textit{J}'}}{(2J'+1)}.
\end{equation}
Eq.~(\ref{collcoeffs}) implies that the collisional processes for each magnetic sublevel are treated equally. Finally, given that we have no knowledge regarding the collisional coefficients between magnetic sublevels (hereafter denoted as $C^{\prime}_{\textit{m}\rightarrow \textit{m}'}$), we assume that $C^{\prime}_{\textit{m}\rightarrow \textit{m}'}$ = $C_{\textit{J,m}\rightarrow \textit{J}',\textit{m}'}$ for all of our tests, unless otherwise stated. However, in our numerical implementation, the collisional coefficients between magnetic sublevels can be scaled as $C^{\prime}_{\textit{m}\rightarrow \textit{m}'}$ = $f_{GK} \times C_{\textit{J,m}\rightarrow \textit{J}',\textit{m}'}$, where $f_{GK}$ is a user-defined variable.

Finally, once the level populations are calculated, we compute the specific intensity for every frequency and position $(l, b)$ on the front face of the simulation (i.e. the ``plane of the sky'') by integrating the radiative-transfer equation for all points ``$k$'' along the line of sight
\begin{equation}\label{rt_eq}
I^{q}_{k+1}=\frac{(e^{-\tau_{k+1}^C}-\zeta^{q})I^{q}_l+\zeta^{q}S_k^{q}+\xi^{q}S_{k+1}^{q}+ \mathcal{S}}{1+\xi^{q}}
\end{equation}
where $\tau^C$ is the optical depth for continuum emission and $\mathcal{S}$ is a function of the source function for dust-continuum emission, and the optical depth and absorption coefficient for continuum (see Eq. 4 from Tritsis et al. 2018). For more details on the dust model used we refer the reader to Tritsis et al. (2018). The quantities $\xi$ and $\zeta$ are defined as
\begin{equation}\label{d}
\xi=\frac{\tau_{k+1}^q}{1+e^{-\tau_{k+1}^q}}
\end{equation}  
\begin{equation}\label{p}
\zeta=\xi(e^{-\tau_{k+1}^q-\tau_{k+1}^C})
\end{equation}
where the optical depth for line emission $\tau^q$ is calculated by integrating the absorption coefficient between points ``$k$'' and ``$k+1$'' and for the normalized profile function we use
\begin{equation}\label{normprof}
\phi = \frac{1}{\Delta \varv_{i, j, k}\sqrt{\pi}} e^{-\big(\frac{\nu - \nu_{J, J'}(1 + \frac{\vec{\hat{w}_{LOS}} \cdot \vec{\varv_{i, j, k}}}{c})}{\Delta \varv_{i, j, k}}\big)^2}
\end{equation}
where $\vec{\hat{w}_{LOS}}$ is the unit vector that defines the line-of-sight (LOS) direction and $\vec{\varv_{i, j, k}}$ is the velocity of grid point ($i, j, k$).

\section{Benchmarking}\label{bench}

For our initial radiative-transfer tests, we use as input a 2D cylindrical, isothermal, non-ideal MHD chemo-dynamical simulation of a collapsing prestellar core presented in Tritsis et al. (2023) (see also Tritsis et al. 2022 for a detailed description of the methodology followed for performing these chemo-dynamical simulations). We use the model with a temperature of $T= 10~\rm{K}$, visual extinction of $A_v =10$, a standard cosmic-ray ionization rate ($\zeta=1.3\times10^{-17} \rm{s^{-1}}$; Caselli et al. 1998) and an initial mass-to-flux ratio (normalized to the critical value; Mouschovias \& Spitzer 1976) of 1/2 (see Table~1 of Tritsis et al. 2023). 

\begin{figure*}
\includegraphics[width=2.075\columnwidth, clip]{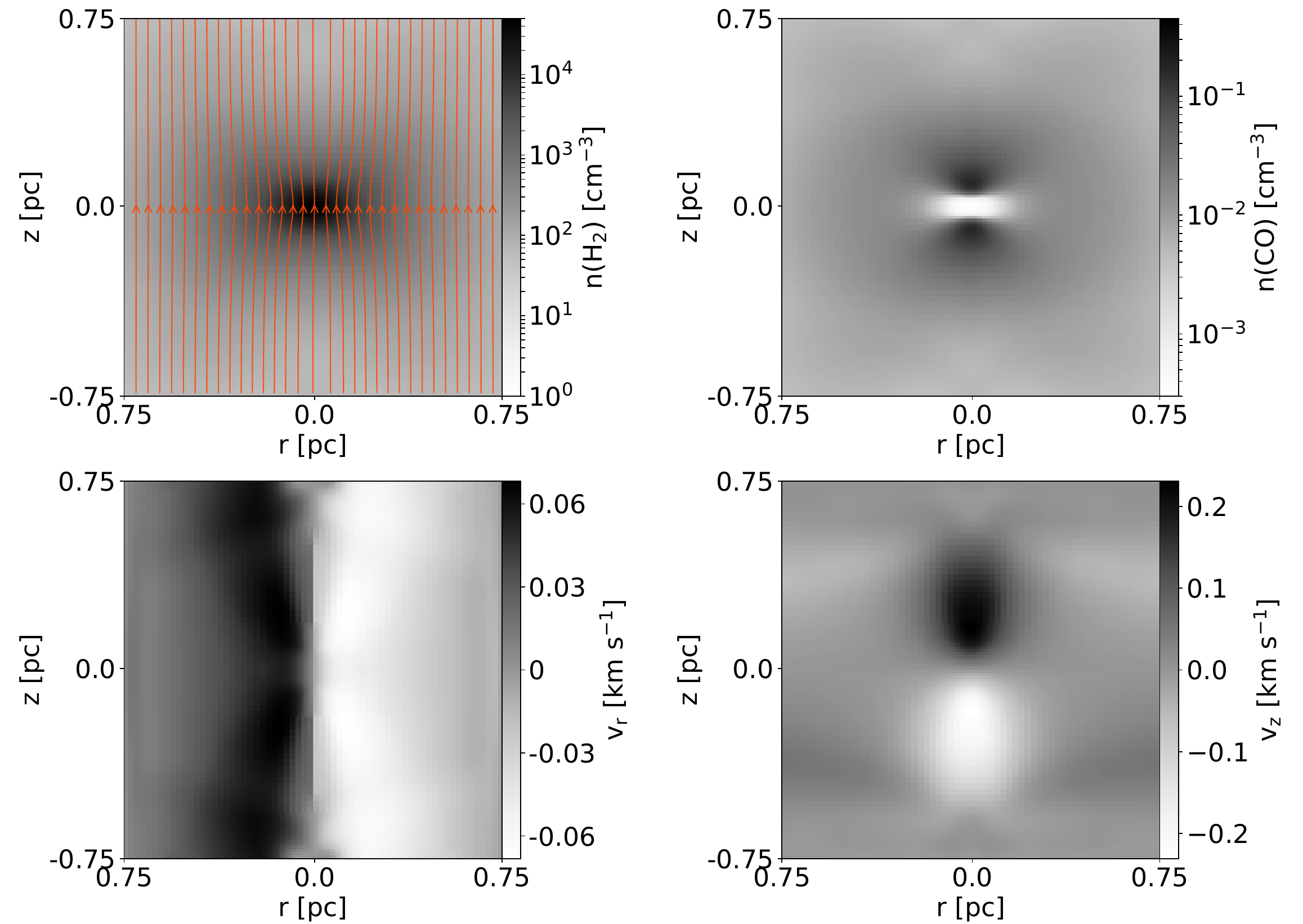}
\caption{In the upper left and right panels we show the $\rm{H_2}$ number density, overlaid with orange streamlines depicting the magnetic field lines, and the CO number density, respectively. In the bottom two panels we show the $r$-velocity component (left) and $z$-velocity component (right).
\label{pyrateinputs}}
\end{figure*}

We post-process this chemo-dynamical simulation when the central density is $n_{\rm{H_2}} = 5\times 10^4~\rm{cm^{-3}}$. In Fig.~\ref{pyrateinputs} we show the $\rm{H_2}$ and CO number densities in the upper left and right panels, respectively. The orange streamlines overlaid on top of the $\rm{H_2}$ number density show the magnetic field lines. The bottom left and right panels in Fig.~\ref{pyrateinputs} show, respectively, the $r$ and $z$ components of the velocity.

\subsection{Two-level molecule under LTE}

\begin{figure*}
\includegraphics[width=2.1\columnwidth, clip]{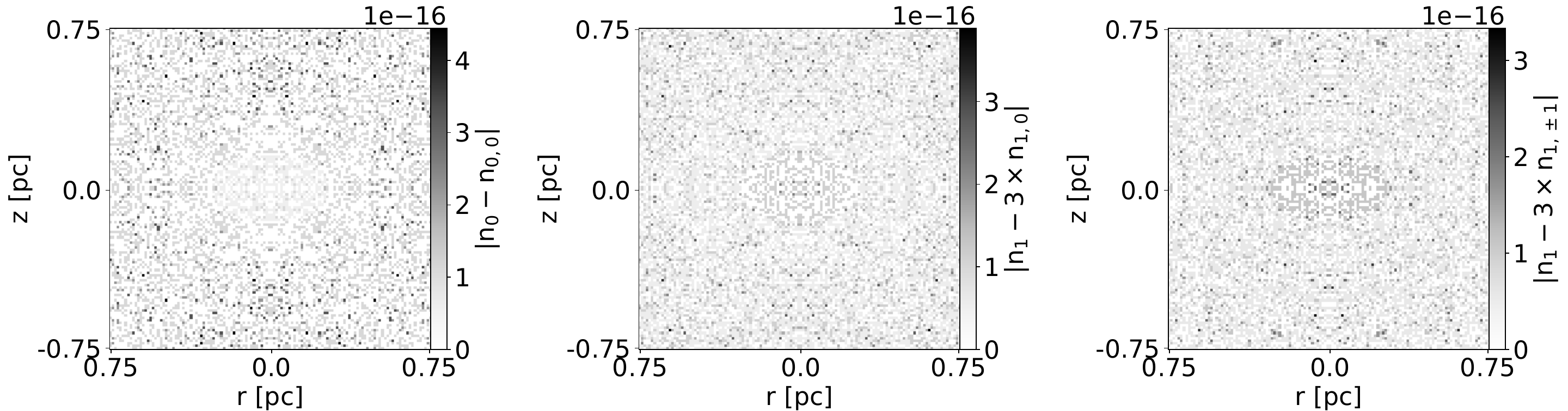}
\caption{Comparison of the level populations throughout the simulated core shown in Fig.~\ref{pyrateinputs}, under LTE conditions. When the level populations are designated without a second subscript to denote the magnetic quantum number all sublevel populations are degenerate. In the opposite case, the magnetic sublevels are explicitly considered in the detailed balance equations. As expected, under LTE conditions, the numerical result is $n_{0, 0} = n_0$ and $n_{1, 0} = n_{1,\pm 1} = \frac{1}{3}n_{1}$, irrespectively of the local physical conditions. Note that the differences are in units of $10^{-16}$.
\label{testTwoLevs}}
\end{figure*}

The aim of our first numerical experiment to test our implementation is to ensure that no spurious linear polarization is present in our calculations in cases where the linear polarization should be zero (i.e. under LTE conditions). Additionally, we want to ensure that in limiting cases our calculations revert back to the ``fiducial'' case where the magnetic-sublevel populations are degenerate. To do so, we compare the level populations when the magnetic sublevels are considered individually with the level populations computed under the fiducial case. For this numerical test, we only consider two energy levels, $J = 0 - 1$, and the contribution from the CMB in Eq.~(\ref{lvgapprox}) is taken into account.

In the left panel of Fig.~\ref{testTwoLevs} we show the residual between the level population of the zeroth rotational level computed when all magnetic-sublevel populations are degenerate, defined as $n_0$ (that is without a second subscript to denote the magnetic quantum number), with the level population of the zeroth rotational level computed when the magnetic sublevels are explicitly considered in the detailed balance equations. As expected, the numerical result is $n_0 = n_{0, 0}$, throughout the simulated core and irrespective of the physical conditions in each cell down to numerical accuracy. In the middle and right panels we show the residuals between the level populations of the three magnetic sublevels of the first rotation level ($n_{1, 0}$ and $n_{1, \pm 1}$, respectively) and the level population of the first rotational level when there is no external magnetic field ($n_1$). The code correctly produces $n_{1, 0} = n_{1,\pm 1} = \frac{1}{3}n_{1}$. Given the fact that the contribution from the CMB is taken into account during the calculation of the level populations, this numerical experiment also demonstrates that isotropic radiation \textit{cannot} lead to unequal populations in the different magnetic sublevels (see \S~\ref{intro} for the conditions required for the GK effect to arise).	

\begin{figure}
\includegraphics[width=1.\columnwidth, clip]{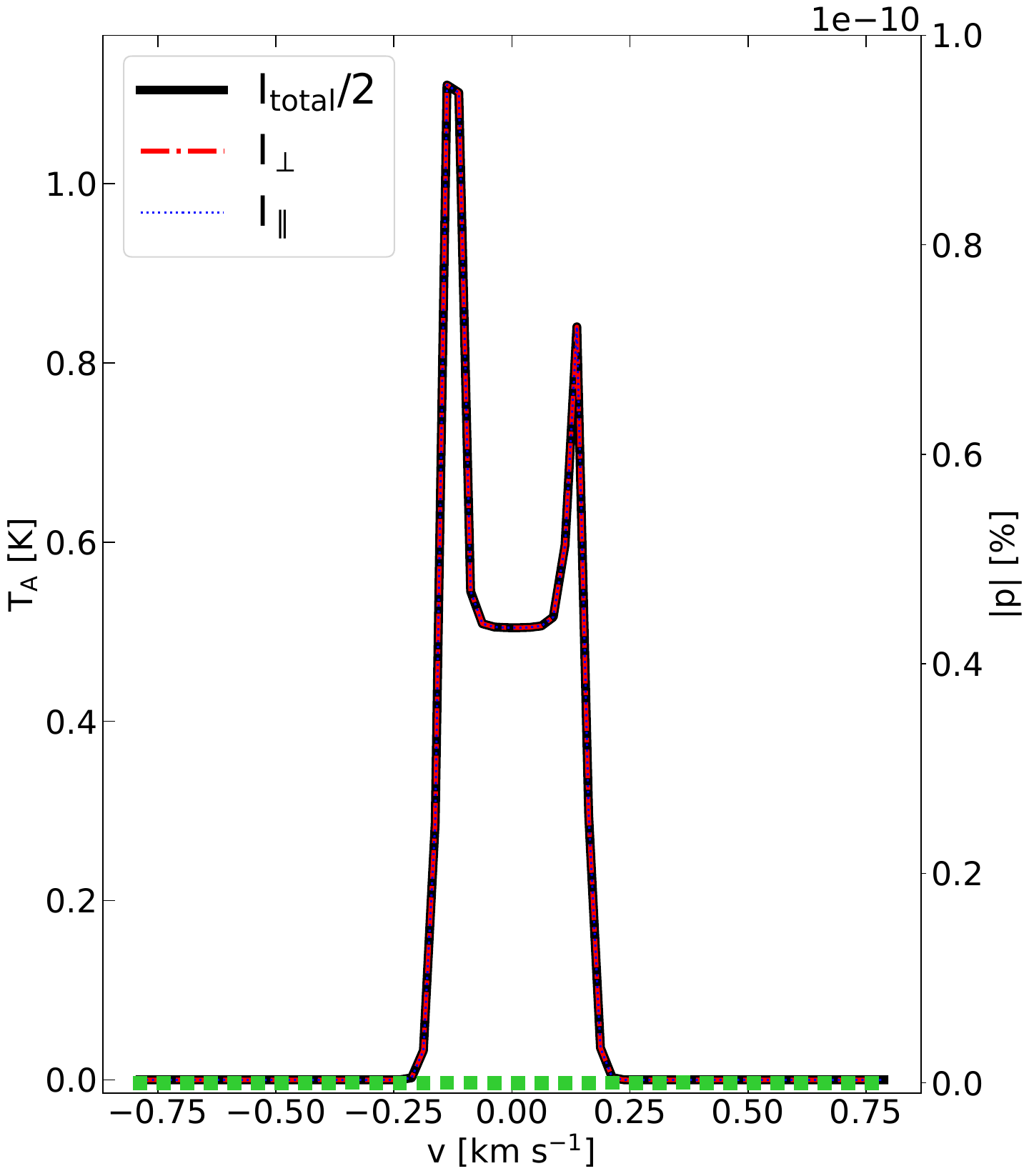}
\caption{Comparison of the spectrum (in antenna temperature units) computed when the magnetic-sublevel populations are degenerate (solid black line) under LTE conditions, with the spectra polarized perpendicular and parallel to the magnetic field (red dashed-dotted and blue dotted lines, respectively), computed when the magnetic sublevels are explicitly considered in the detailed balance equations. All spectra are from the central cell in our simulation box ($r=z=0$). The green points show the polarization fraction (right $y$ axis). As expected, under LTE conditions, $I_\perp = I_\parallel = I_{total}/2$ and the polarization fraction is zero to numerical accuracy (note that the polarization fraction is measured in units of $10^{-10}$).
\label{testTwoLevsLinePol}}
\end{figure}

In Fig.~\ref{testTwoLevsLinePol} we show spectra in antenna temperature units from the center of the core computed for each of the two cases described above. With the solid black line we show the spectrum when the magnetic-sublevel populations are degenerate. With the red dashed-dotted and blue dotted lines we show respectively the spectra polarized perpendicular and parallel to the magnetic field. As expected, under LTE conditions, $I_{total}/2 = I_{\perp} = I_{\parallel}$. Additionally, the polarization fraction defined as
\begin{equation}\label{polFrac}
p = \frac{I_\perp - I_\parallel}{I_\perp + I_\parallel - B}
\end{equation}
is zero to numerical accuracy (see green squares in Fig.~\ref{testTwoLevsLinePol} corresponding to the right $y$ axis).

\subsection{Multilevel molecule under LTE}

Here, we extend the calculations performed in the previous section for a multilevel molecule with four rotational energy levels ($J = 0-3$). For four rotational levels, we need to compute the level populations for ten magnetic sublevels. Similarly to the previous section, the contribution from the CMB is considered in our calculations of the level populations.

\begin{figure*}
\includegraphics[width=2.05\columnwidth, clip]{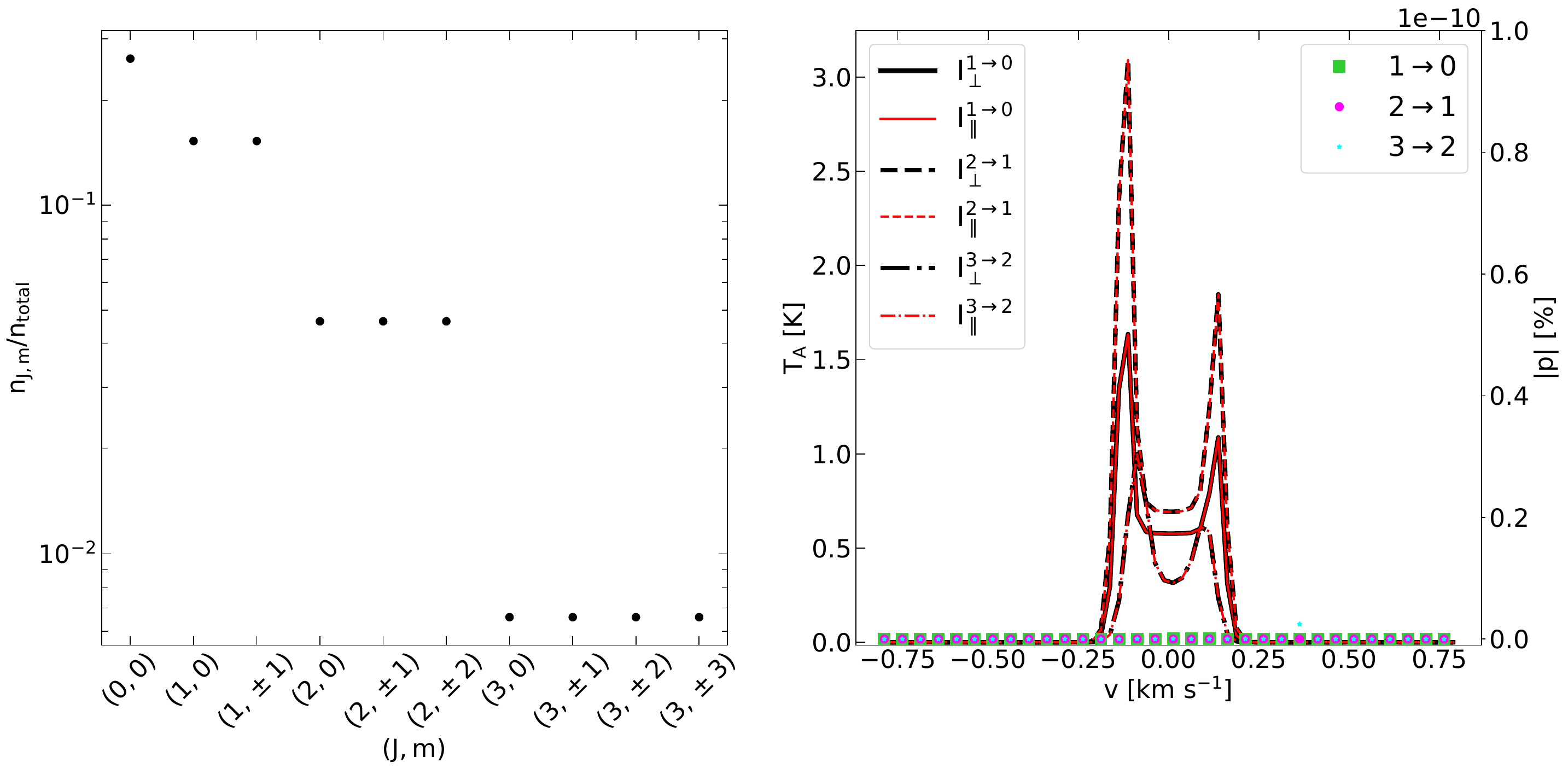}
\caption{Left panel: population fraction of the different magnetic sublevels at the center of the core under LTE conditions. Similarly to the two-level approximation, the level populations of the magnetic sublevels of the same rotational level are equal under LTE conditions. Right panel: Spectra of the different transitions towards the center of the cloud (i.e. $r=z=0$) and the corresponding polarization fraction (right $y$ axis; see legends for the definition of the different linestyles and colors). Once again, there is no spurious polarization fraction for any pair of rotational transitions.
\label{testMultiLTE}}
\end{figure*}

In the left panel of Fig.~\ref{testMultiLTE} we show the population fraction of the different magnetic sublevels at the center of the core. As expected, under LTE conditions, the magnetic sublevels of the same rotational levels have equal level populations. In the right panel of Fig.~\ref{testMultiLTE} we show the spectra from the CO $J = 1\rightarrow 0$ (solid lines), $J = 2\rightarrow 1$ (dashed lines) and $J = 3\rightarrow 2$ (dashed-dotted) transitions. With the black lines we show the intensity polarized perpendicular to the magnetic field and with the red lines we show the intensity polarized parallel to the magnetic field. Here, we also show the polarization fraction from each transition, which once again is zero to numerical accuracy.

\subsection{Two-level molecule under non-LTE}\label{AnalyticalComp}

Now that we have established that no spurious linear polarization is present in our implementation, we proceed to test whether the numerical results for the polarization fraction under non-LTE conditions are in agreement with theoretical expectations. To do so, we compare the numerical result for the fractional polarization at the rest frequency of the line ($\nu_{J, J'}$) with the analytical results of Kylafis (1983) for the polarization fraction as a function of the ``mean'' optical depth. For this numerical test, we match the physical conditions as well as the Einstein $A$ and collisional coefficients adopted by Kylafis (1983). Specifically, the only non-zero velocity component is taken to be along the magnetic-field direction with the velocity gradient being set equal to $\Lambda = 10^{-11}~\rm{s^{-1}}$. In the directions perpendicular to the field the cloud is taken to be infinite or, in other words, the optical depths are set to infinity. The line-of-sight direction is perpendicular to the magnetic field such that the amount of polarization is maximum for the specific physical conditions under consideration. The collisional and Einstein $A$ coefficients are set equal to $9.4\times 10^{12}~\rm{cm^3~s^{-1}}$ and $1.8\times 10^{-7}~\rm{s^{-1}}$, respectively and the $\rm{H_2}$ number density is set equal to $1.9\times 10^4~\rm{cm^{-3}}$, such that $Cn_{\rm{H_2}}/A$ is equal to one. Finally the temperature is set equal to 30 K everywhere in our simulation grid. For this test, the contribution from the CMB in Eq.~(\ref{lvgapprox}) is ignored, as in the study by Kylafis (1983).

\begin{figure}
\includegraphics[width=1.\columnwidth, clip]{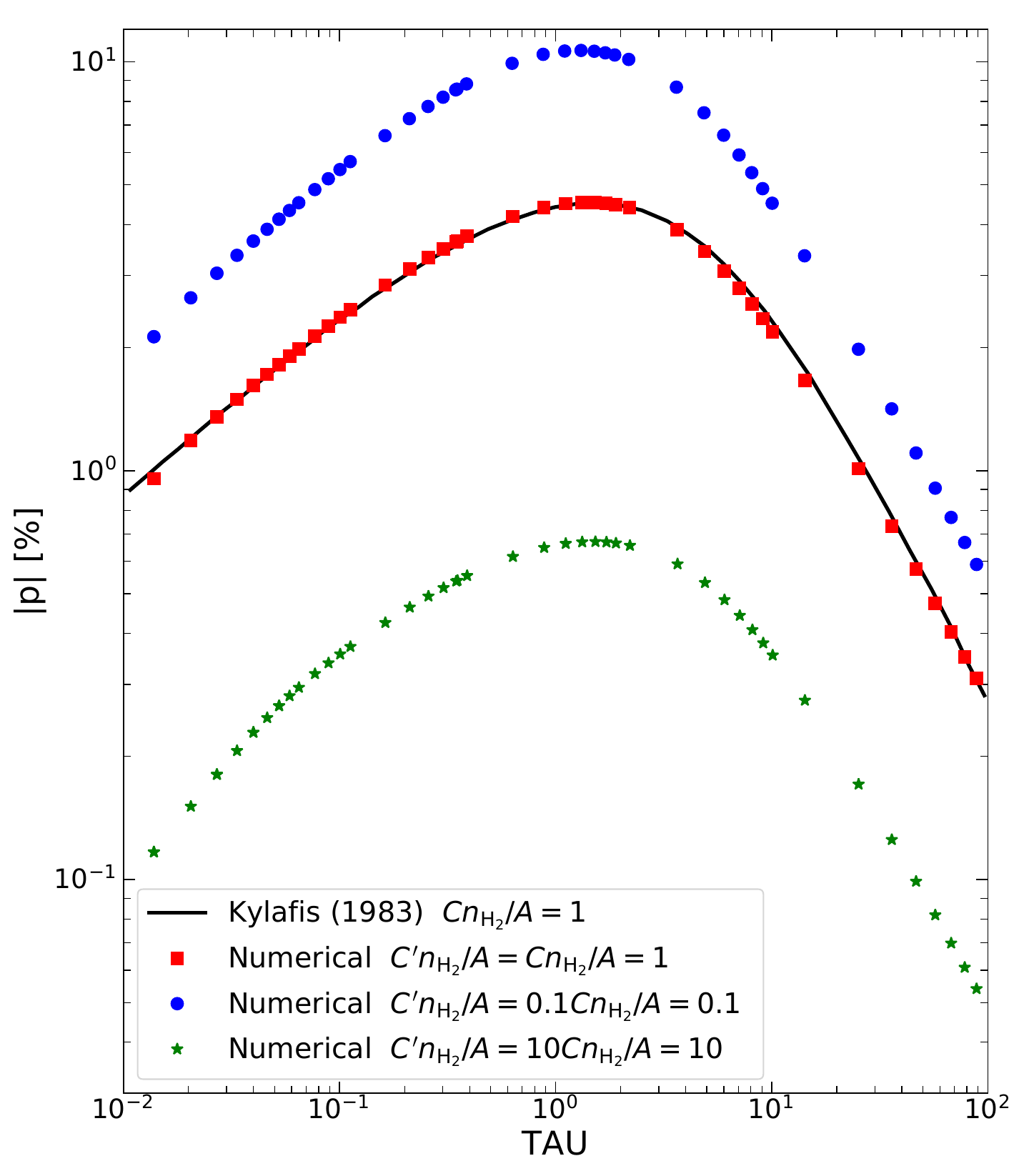}
\caption{Polarization fraction as a function of the ``mean" optical depth (see \S~\ref{AnalyticalComp}). The black solid line shows the analytical results by Kylafis (1983) for $Cn_{\rm{H_2}}/A = 1$ and red squares show our numerical results for the same value of $Cn_{\rm{H_2}}/A$ and with $C^{\prime}_{\textit{m}\rightarrow \textit{m}'}$ = $C_{\textit{J,m}\rightarrow \textit{J}',\textit{m}'}$. Green stars and blue circles show our numerical results when we alter the value of $C^{\prime}_{\textit{m}\rightarrow \textit{m}'}$ a factor of ten above and below $C_{\textit{J,m}\rightarrow \textit{J}',\textit{m}'}$.
\label{PolfTAU}}
\end{figure}

In Fig.~\ref{PolfTAU} we show the polarization fraction at the rest frequency of the line as a function of the ``mean'' optical depth $\rm{TAU}$, defined in Appendix B of Kylafis (1983) as
\begin{equation}\label{meanTau}
\frac{1}{\rm{TAU}} = \int \frac{d\Omega}{4\pi}\frac{1}{\tau} = \frac{1}{3\tau_0},
\end{equation}
where $\tau$ is the optical depth as a function of direction and
\begin{equation}\label{Tau0}
\tau_0 = \frac{3}{8\pi}\Big(\frac{c}{\nu_{J, J'}}\Big)^3 A_{\textit{J,m}\rightarrow \textit{J}',\textit{m}'} \frac{n_{0} - \frac{2n_{1,\pm 1} + n_{1,0}}{3}}{\Lambda}.
\end{equation}

With red squares we plot our numerical results for $Cn_{\rm{H_2}}/A = 1$ and the solid black line shows the analytical results for the same value of $Cn_{\rm{H_2}}/A$. As evident, the numerical results follow very well the analytical solution across all values of the ``mean'' optical depth. Finally, with blue circles and green stars in Fig.~\ref{PolfTAU} we show our numerical results when $Cn_{\rm{H_2}}/A = 1$ but with $C^{\prime}_{\textit{m}\rightarrow \textit{m}'}$ = $0.1 \times C_{\textit{J,m}\rightarrow \textit{J}',\textit{m}'}$ and $C^{\prime}_{\textit{m}\rightarrow \textit{m}'}$ = $10 \times C_{\textit{J,m}\rightarrow \textit{J}',\textit{m}'}$, respectively. Increasing/decreasing the collisional coefficient between magnetic sublevels by a factor of ten leads to a significant decrease/increase in the polarization fraction. This is to be expected, as collisions tend to populate different sublevels equally and therefore the polarization fraction is inversely correlated to $C^{\prime}_{\textit{m}\rightarrow \textit{m}'}$. However, as it is evident from Fig~\ref{PolfTAU}, the effect is non-linear. That is, decreasing the value of $C^{\prime}_{\textit{m}\rightarrow \textit{m}'}$ leads to an increase in the polarization fraction by a factor of $\sim$3, whereas increasing the value of $C^{\prime}_{\textit{m}\rightarrow \textit{m}'}$ by the same factor has a more dramatic effect and leads to a decrease in the polarization fraction by almost an order of magnitude.

\subsection{Multilevel molecule under non-LTE}

We now explore how the polarization fraction is affected when multiple rotational levels are considered in our calculations. To this end, we consider a total of four rotational levels ($J = 0 -3$). The underlying physical model for the calculations presented here is identical to the one considered in the previous section. However, we now use the collisional and Einstein coefficients from the \textsc{LAMBDA} database (Sch{\"o}ier et al. 2005). The value of the $\rm{H_2}$ number density is set equal to $ 2.1\times 10^2/10^3/10^4~\rm{cm^{-3}}$, such that $Cn_{\rm{H_2}}/A$ is 0.1, 1 and 10, respectively. Finally, the value of $C^{\prime}_{\textit{m}\rightarrow \textit{m}'}$ is set equal to $C_{\textit{J,m}\rightarrow \textit{J}',\textit{m}'}$. As in the previous numerical experiment the contribution from the CMB is not taken into account when computing the population densities of the different sublevels.

\begin{figure}
\includegraphics[width=1.\columnwidth, clip]{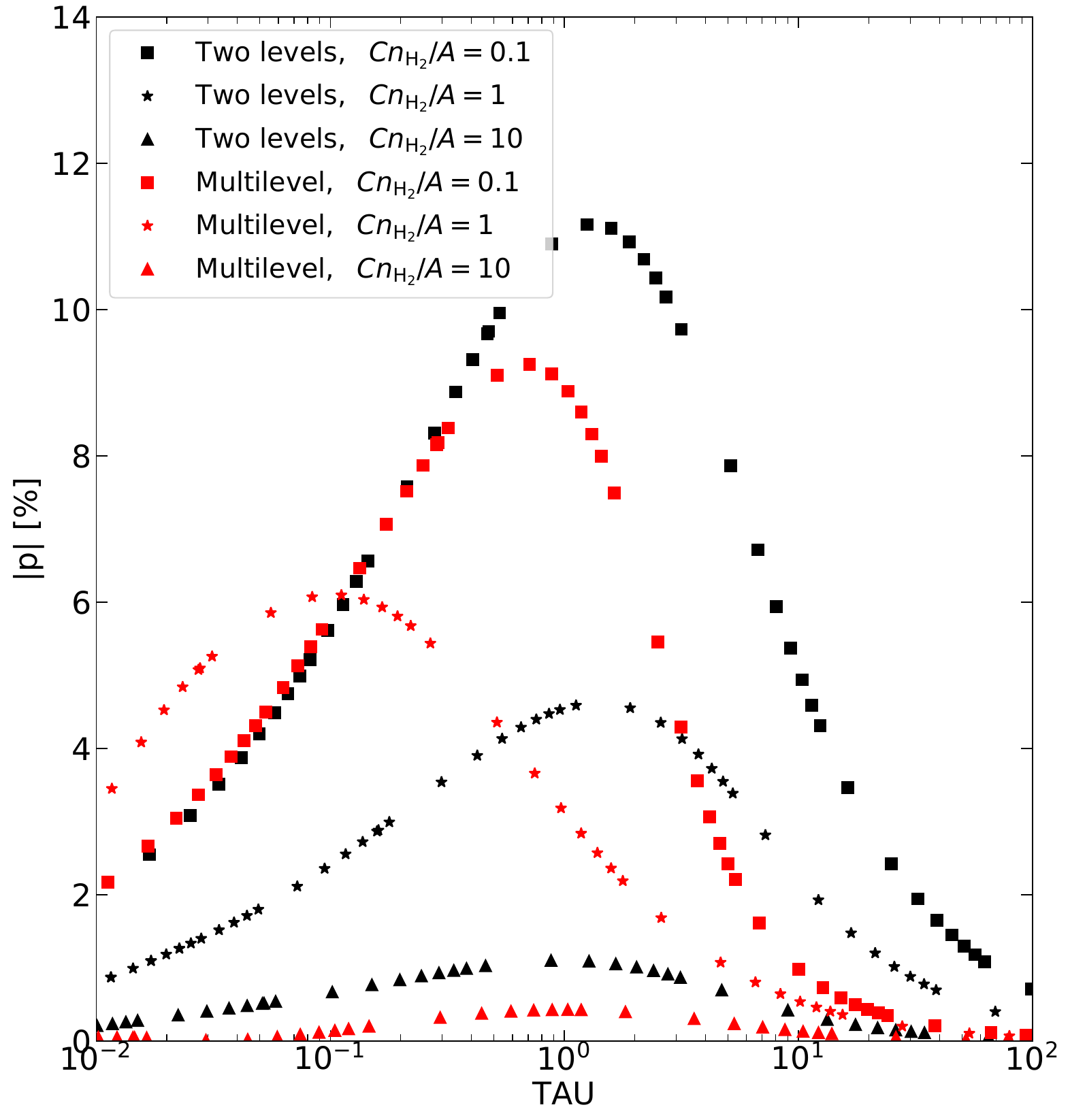}
\caption{Polarization fraction as a function of the ``mean" optical depth for different values of $Cn_{\rm{H_2}}/A$ for the multilevel and two-level cases (red and black points, respectively). With triangles, stars and squares we show the polarization fraction for $Cn_{\rm{H_2}}/A = $ 10, 1 and 0.1, respectively.
\label{multileveltest}}
\end{figure}

In Fig.~\ref{multileveltest} we show the polarization fraction from CO $J = 1\rightarrow 0$ transition in the two-level and multilevel cases for the three different values of $Cn_{\rm{H_2}}/A$. The black points show our calculations for the two-level case and the red points when considering multiple rotational levels. With triangles we show our results for $Cn_{\rm{H_2}}/A = 1$, with stars we show our results for $Cn_{\rm{H_2}}/A = 1$ and finally with the squares we show our results for $Cn_{\rm{H_2}}/A = 0.1$. As it is evident from Fig.~\ref{multileveltest}, considering more rotational levels has a non-trivial effect in the fractional polarization. Specifically, when considering multiple levels, the peak in polarization fraction is observed at smaller ``mean'' optical depths. This is mostly evident in the case when $Cn_{\rm{H_2}}/A = 1$ whereas for $Cn_{\rm{H_2}}/A = $ 10 and 0.1 the peak in polarization fraction is for ``mean'' optical depths close to unity, as in the two-level case. A very similar trend was obtained by Deguchi \& Watson (1984) (see their Figure 4), although the exact value of the polarization fraction predicted here differs from theirs for the same values of $Cn_{\rm{H_2}}/A$.

We should note however that there are a number of differences between the numerical calculations presented here and the calculations by Deguchi \& Watson (1984). Firstly, as noted in \S~\ref{numer}, we use Eq.~(\ref{EinAcorrect}) instead of Eq.~(\ref{EinAwrong}) for computing the value of the Einstein $A$ coefficient between all transitions $(J, m)\rightarrow(J^\prime, m^\prime)$. Additionally, Deguchi \& Watson (1984) ignore collisions between magnetic sublevels whereas such interactions are taken into account for the results presented in Fig.~\ref{multileveltest}. Finally, for the collisional coefficients $C_{\textit{J,m}\rightarrow \textit{J}',\textit{m}'}$, Deguchi \& Watson (1984) adopt their values from Green \& Chapman (1978), which differ from the values used here by $\sim$30\%. Unfortunately, however, a one-to-one comparison between the different implementations is not possible, as Deguchi \& Watson (1984) do not explicitly quote the values they use for the Einstein $A_{\textit{J}\rightarrow \textit{J}'}$ coefficients\footnote{The detailed balance equations (see Eq.~\ref{seeEq2}) can be divided by $A_{\textit{J, m}\rightarrow \textit{J}', \textit{m}'}$ such that only remaining factor is $Cn_{\rm{H_2}}/A$. However, $A_{\textit{J, m}\rightarrow \textit{J}', \textit{m}'}$ still explicitly appears in the calculation of the absorption coefficient (Eq.~\ref{absorp}) which is then used to compute $S_{J, J^\prime}^\perp$ and $S_{J, J^\prime}^\parallel$ (Eqs.~\ref{sourcePepr} \& \ref{sourceParal}). Therefore, the values of Einstein $A_{\textit{J}\rightarrow \textit{J}'}$ coefficients need to be known explicitly to perform a one-to-one comparison.}.


\section{The GK effect in a prestellar core}\label{GKrealistic}

In this section, we present radiative-transfer simulations of the GK effect for $\rm{CO}$ under non-LTE conditions for the physical model shown in Fig.~\ref{pyrateinputs}, considering four rotational energy levels. The use of a non-ideal MHD chemodynamical simulation enables us to have more realistic physical and chemical conditions for our radiative-transfer calculations. The contribution from the CMB is taken into account when computing the population densities and $C^{\prime}_{\textit{m}\rightarrow \textit{m}'}$ is set equal to $C_{\textit{J,m}\rightarrow \textit{J}',\textit{m}'}$. Finally, the core is observed edge-on, such that the mean component of the magnetic field is perpendicular to the line of sight and the fractional polarization is maximum.

\begin{figure*}
\includegraphics[width=2.\columnwidth, clip]{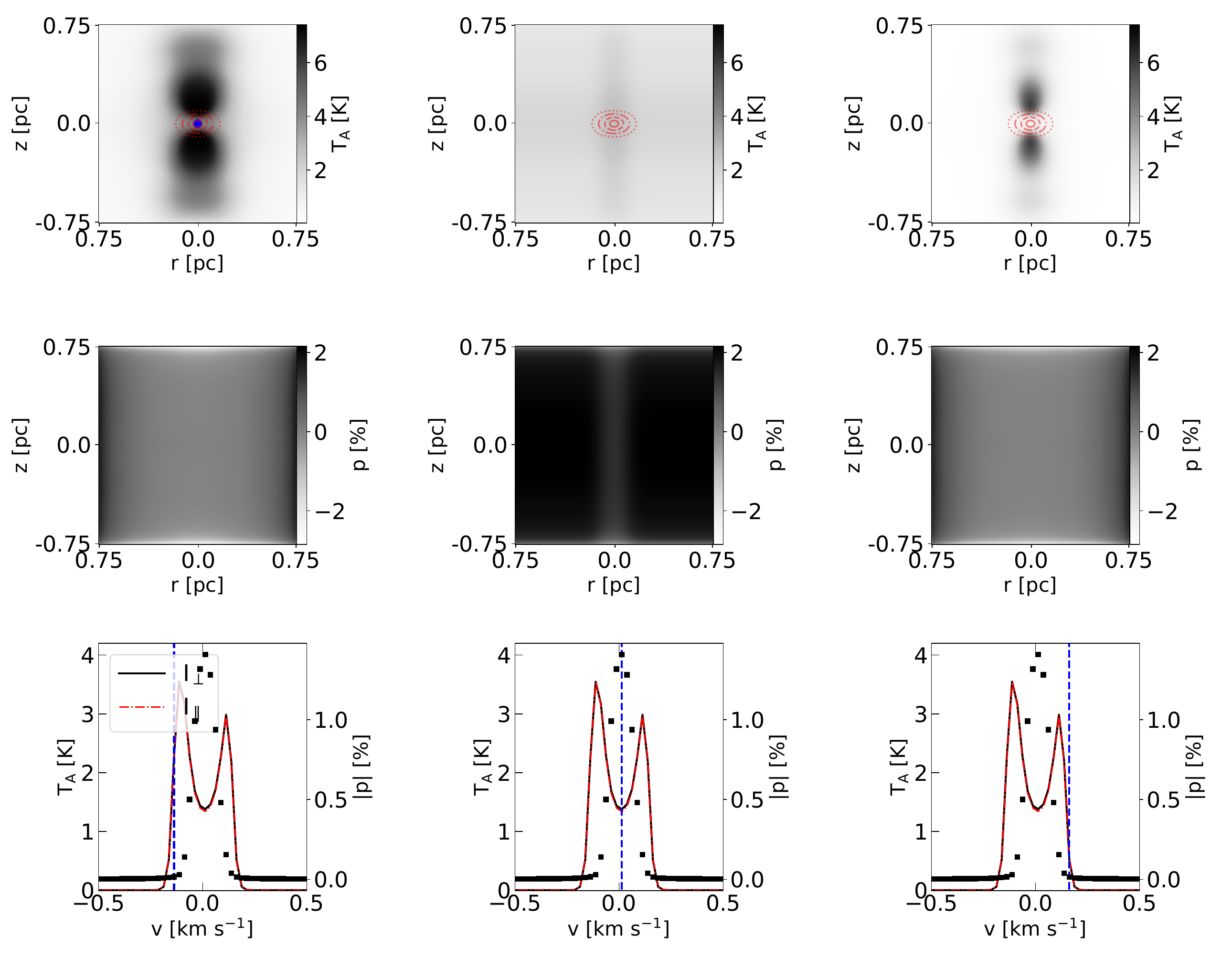}
\caption{Results from our radiative-transfer simulation of the GK effect for the $J = 1\rightarrow 0$ transition for the chemo-dynamical simulation shown in Fig.~\ref{pyrateinputs}. Upper row: Slices through the simulated Position-Position-Velocity (PPV) data cube for the velocities marked with the blue dashed-lines in the lower row where we additionally show spectra at the center of the core. In each panel we have overplotted the actual density structure of the core with red contours. Middle row: Fractional polarization in each velocity slice. Bottom row: Spectra towards the middle of the core (see blue point in the upper left panel) together with the polarization fraction (black squares) for this location in the cloud. With the black line we show the radiation polarized perpendicular to the magnetic field and with the red line we show the radiation polarized parallel to the magnetic field. For the physical conditions of the cloud shown in Fig.~\ref{pyrateinputs} the maximum fractional polarization is observed for the rest frequency of the line (middle panel in the second row).
\label{nonIsim10}}
\end{figure*}

\begin{figure*}
\includegraphics[width=2.\columnwidth, clip]{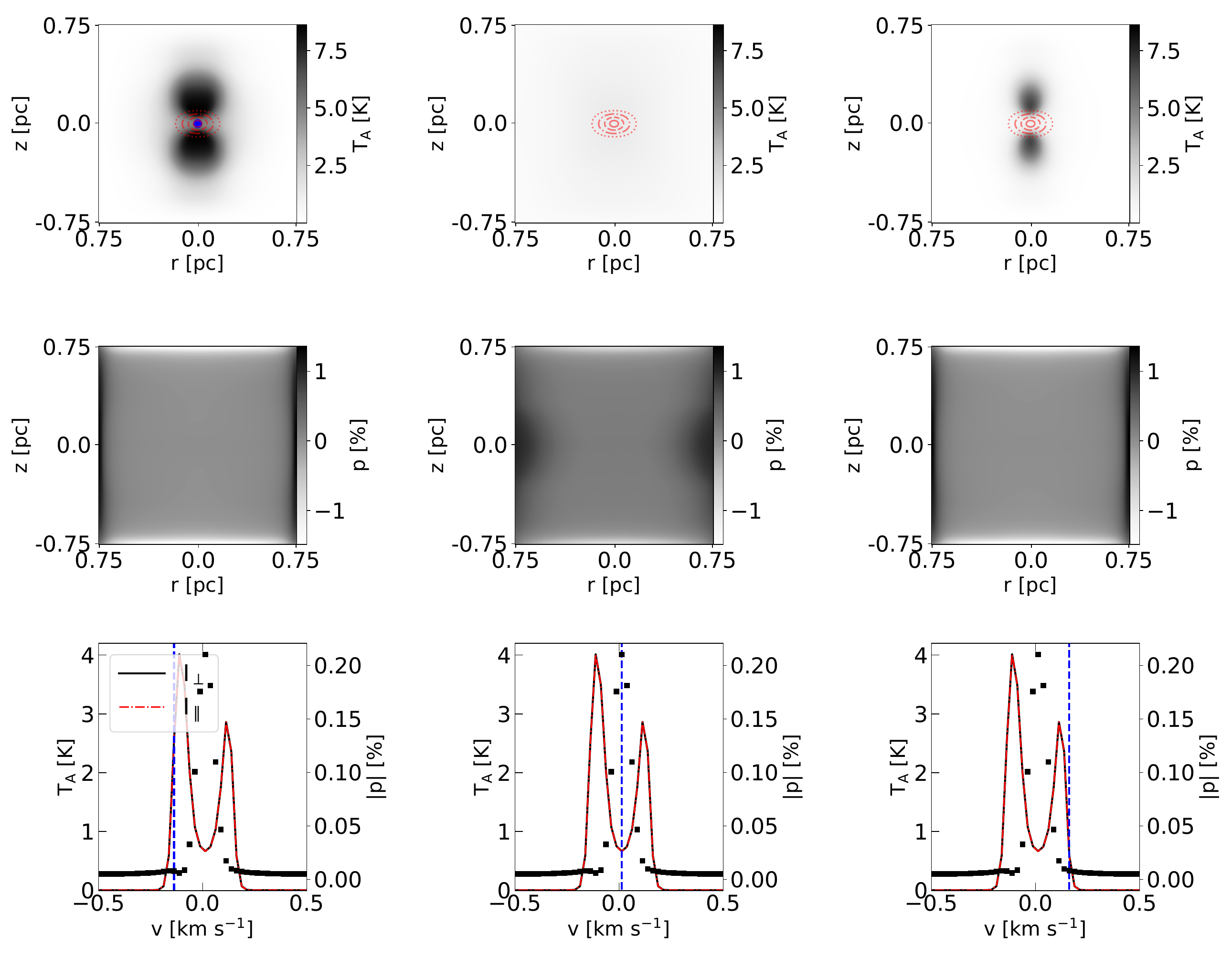}
\caption{Same as Fig.~\ref{nonIsim10} but for the $J = 3\rightarrow 2$ transition.  
\label{nonIsim32}}
\end{figure*}

In Figs.~\ref{nonIsim10} \& ~\ref{nonIsim32} we present our results from our numerical calculations for the $J = 1\rightarrow 0$ and $J = 3\rightarrow 2$ transitions. Results for the $J = 2\rightarrow 1$ transition are qualitatively very similar to the $J = 1\rightarrow 0$ transition and are therefore not shown here. In the upper row we show (in units of antenna temperature) the intensity of the line ($I_{total} = I_{\perp} + I_{\parallel}$) in three different slices through our mock Position-Position-Velocity (PPV) data cube. The red contours show the actual density structure of the core (see Fig.~\ref{pyrateinputs}). The velocity of these slices is marked with the blue dashed lines in the bottom row, where we additionally show a spectrum through the center of the core. Finally, in the middle row we show the fractional polarization in the entire core for each velocity slice.

As expected, the fractional polarization is maximum at the rest frequency for both the $J = 1\rightarrow 0$ and the $J = 3\rightarrow 2$ transitions. Specifically, for the $J = 1\rightarrow 0$ transition the polarization fraction is of the order of $\sim$2\%, whereas for the $J = 3\rightarrow 2$ transition the polarization fraction towards the middle of the cloud is an order of magnitude less. Given the Einstein $A$ and collisional coefficients, the factor $Cn_{\rm{H_2}}/A$ for the $J = 3\rightarrow 2$ transition remains close to unity even in the inner regions of the cloud as opposed to the $J = 1\rightarrow 0$ transition where it reaches a maximum value of $\sim$24. Hence, the fact that the polarization fraction is higher for the $J = 1\rightarrow 0$ might at first seem counter-intuitive. However, the $J = 3\rightarrow 2$ transition is more optically thick with typical values of the optical depth being at an excess of 200. Therefore, a lower fractional polarization is to be expected especially considering that in the multilevel case the peak in polarization fraction is found at slightly smaller optical depths than the two-level case (see Fig.~\ref{multileveltest}). As a result, in neither of the two transitions the polarized radiation comes from the inner regions of the core, but for (mostly) different physical reasons, which are further explained below.

Spatially, a small drop in polarization fraction is observed towards the axis of symmetry of the cloud for the $J = 1\rightarrow 0$ transition (from 2.2\% to 1.3\%; see middle panel in the second row in Fig.~\ref{nonIsim10}). This drop in polarization fraction is a combined effect of two factors. Firstly, the $\rm{H_2}$ number density increases near the axis of symmetry and consequently $Cn_{\rm{H_2}}/A$ increases, leading to a decrease in polarization fraction. Secondly, and most importantly, the optical depth increases near the axis of symmetry leading to a further decrease in polarization fraction. This can be intuitively understood given the fact that photons emitted from regions of the cloud close to the axis of symmetry need to travel a longer distance. In contrast to the $J = 1\rightarrow 0$ transition, the polarization fraction in the $J = 3\rightarrow 2$ transition remains relatively uniform in the inner regions of the cloud and only increases to observationally-detectable values towards the outer regions of the core, where both $Cn_{\rm{H_2}}/A$ and the optical depth are low.


At the rest frequency of the $J = 1\rightarrow 0$ transition the polarization fraction is positive through the entire core and only becomes negative in the outer regions of the core for frequencies close to $\sim$one thermal linewidth away from the rest frequency (left and right panels in the second row of Fig.~\ref{nonIsim10}). This implies that, at the rest frequency of this transition, the polarization is perpendicular to the magnetic field, but changes to being parallel in the outer regions of the core for other velocity slices. The same is largely true for the $J = 3\rightarrow 2$ transition. However, in this case the polarization becomes parallel to the magnetic field (negative polarization fraction) even for the rest frequency of the line and for $|z|\gtrsim$0.65 pc (middle panel in the second row of Fig.~\ref{nonIsim32}). Such differences in the polarization direction occur because the factor $\tau^{\perp}_{J', J} - \tau^{\parallel}_{J', J}$ changes sign in these regions of the cloud. While changes in polarization direction between different transitions of the same molecule (and for the same region of the cloud) have been previously pointed out in the literature (e.g. Cortes et al. 2005), not much attention has been paid to variations in the polarization direction for the same transition. However, such variations have potentially been observed. For instance, Lai et al. (2003) found that $\rm{CO}$ and dust polarization vectors where aligned in one region of DR21(OH), whereas they were perpendicular in another region (see their Fig. 1). Such variations in the polarization direction in different, and the same transition(s), in combination with dust polarization observations and numerical simulations (e.g. Bino et al. 2022) could potentially be used to probe the velocity component in the plane of the sky.

\section{Summary and Conclusions}\label{discuss}

We implemented a multilevel treatment of the GK effect in the non-LTE line radiative-transfer code \textsc{PyRaTE}, where we individually treat the different modes of polarized radiation. We tested our implementation for various limiting cases and compared our numerical calculations against analytical results. Firstly, we confirmed that under LTE conditions the fractional polarization is zero to numerical accuracy, in both the two-level and multilevel cases. Additionally, we confirmed that even when the magnetic sublevels are explicitly considered, under LTE conditions, our results are identical to the case where the magnetic-sublevel populations are degenerate. We then compared our numerical calculations for the fractional polarization as a function of the ``mean'' optical depth against analytical results and found an excellent agreement.

Finally, we presented radiative-transfer simulations of the GK effect in $\rm{CO}$ ($J = 1\rightarrow 0$ and $J = 3\rightarrow 2$ transitions) using as input a chemo-dynamical, non-ideal MHD simulation of a prestellar core when the central number density of the cloud is $n_{\rm{H_2}} = 5\times 10^4~\rm{cm^{-3}}$. At the rest frequency of the transitions, we found a relatively uniform polarization fraction throughout the inner regions of the cloud of the order of 2\% and 0.2\%, respectively.

With our new implementation, we can provide observationally-testable predictions for the polarization fraction for any set of given physical parameters. Such predictions include the variation of the polarization fraction both spatially within an interstellar cloud and as a function of velocity. This synergy between simulations, observations can open new pathways for studying magnetic fields in various stages during the star-formation process as well as potentially revealing the, previously inaccessible, plane-of-sky component of the velocity field during the early stages in the star-formation process. The code is freely available to download at \hyperlink{here}{https://github.com/ArisTr/PyRaTE.git}.

\begin{acknowledgements}

We thank the anonymous referee for suggestions that improved this manuscript. A. Tritsis acknowledges support by the Ambizione grant no. PZ00P2\_202199 of the Swiss National Science Foundation (SNSF). The software used in this work was in part developed by the DOE NNSA-ASC OASCR Flash Center at the University of Chicago. We also acknowledge use of the following software: \textsc{Matplotlib} (Hunter 2007), \textsc{Numpy} (Harris et al. 2020), \textsc{Scipy} (Virtanen et al. 2020) and the \textsc{yt} analysis toolkit (Turk et al. 2011).

\end{acknowledgements}

%

\begin{thebibliography}{}

\bibitem[Barnes et al.(2023)]{2023ApJ...945...34B} Barnes, P.~J., Ryder, S.~D., Novak, G., et al.\ 2023, \apj, 945, 34. doi:10.3847/1538-4357/acac27

\bibitem[Bino et al.(2022)]{2022ApJ...936...29B} Bino, G., Basu, S., Machida, M.~N., et al.\ 2022, \apj, 936, 29. doi:10.3847/1538-4357/ac7c0f

\bibitem[Brinch \& Hogerheijde(2010)]{2010A&A...523A..25B} Brinch, C. \& Hogerheijde, M.~R.\ 2010, \aap, 523, A25. doi:10.1051/0004-6361/201015333

\bibitem[Caselli et al.(1998)]{1998ApJ...499..234C} Caselli, P., Walmsley, C.~M., Terzieva, R., et al.\ 1998, \apj, 499, 234. doi:10.1086/305624

\bibitem[Castor(1970)]{1970MNRAS.149..111C} Castor J.~I., 1970, MNRAS, 149, 111. doi:10.1093/mnras/149.2.111

\bibitem[Chandrasekhar \& Fermi(1953)]{1953ApJ...118..113C} Chandrasekhar, S. \& Fermi, E.\ 1953, \apj, 118, 113. doi:10.1086/145731

\bibitem[Crutcher et al.(2009)]{2009ApJ...692..844C} Crutcher, R.~M., Hakobian, N., \& Troland, T.~H.\ 2009, \apj, 692, 844. doi:10.1088/0004-637X/692/1/844

\bibitem[Cortes et al.(2005)]{2005ApJ...628..780C} Cortes, P.~C., Crutcher, R.~M., \& Watson, W.~D.\ 2005, \apj, 628, 780. doi:10.1086/430815

\bibitem[Cort{\'e}s et al.(2021)]{2021ApJ...923..204C} Cort{\'e}s, P.~C., Sanhueza, P., Houde, M., et al.\ 2021, \apj, 923, 204. doi:10.3847/1538-4357/ac28a1

\bibitem[Davis (1951)]{1951Phys.Rev....81...890} Davis, L.\ 1951, Phys. Rev., 81, 890 

\bibitem[Deguchi \& Watson(1984)]{1984ApJ...285..126D} Deguchi, S. \& Watson, W.~D.\ 1984, \apj, 285, 126. doi:10.1086/162483

\bibitem[Deguchi \& Watson(1990)]{1990ApJ...354..649D} Deguchi, S. \& Watson, W.~D.\ 1990, \apj, 354, 649. doi:10.1086/168722

\bibitem[de Jong et al.(1980)]{1980A&A....91...68D} de Jong, T., Boland, W., \& Dalgarno, A.\ 1980, \aap, 91, 68

\bibitem[Falgarone et al.(2008)]{2008A&A...487..247F} Falgarone, E., Troland, T.~H., Crutcher, R.~M., et al.\ 2008, \aap, 487, 247. doi:10.1051/0004-6361:200809577

\bibitem[Forbrich et al.(2008)]{2008A&A...492..757F} Forbrich, J., Wiesemeyer, H., Thum, C., et al.\ 2008, \aap, 492, 757. doi:10.1051/0004-6361:200811056

\bibitem[Girart et al.(2004)]{2004Ap&SS.292..119G} Girart, J.~M., Greaves, J.~S., Crutcher, R.~M., et al.\ 2004, \apss, 292, 119. doi:10.1023/B:ASTR.0000045007.35868.17

\bibitem[Goldreich et al.(1973)]{1973ApJ...179..111G} Goldreich, P., Keeley, D.~A., \& Kwan, J.~Y.\ 1973, \apj, 179, 111. doi:10.1086/151852

\bibitem[Goldreich \& Kylafis(1981)]{1981ApJ...243L..75G} Goldreich, P. \& Kylafis, N.~D.\ 1981, \apjl, 243, L75. doi:10.1086/183446

\bibitem[Goldreich \& Kylafis(1982)]{1982ApJ...253..606G} Goldreich, P. \& Kylafis, N.~D.\ 1982, \apj, 253, 606. doi:10.1086/159663

\bibitem[Green \& Chapman(1978)]{1978ApJS...37..169G} Green, S. \& Chapman, S.\ 1978, \apjs, 37, 169. doi:10.1086/190523

\bibitem[Harris et al.(2020)]{2020Nat.585..357} Harris, C.~R., Millman, K.~J., van der Walt, S.~J. et al.\ 2020 Nature 585, 357–362. doi: 0.1038/s41586-020-2649-2

\bibitem[Houde et al.(2022)]{2022MNRAS.511..295H} Houde, M., Lankhaar, B., Rajabi, F., et al.\ 2022, \mnras, 511, 295. doi:10.1093/mnras/stab3806

\bibitem[Huang et al.(2020)]{2020ApJ...899..152H} Huang, K.-Y., Kemball, A.~J., Vlemmings, W.~H.~T., et al.\ 2020, \apj, 899, 152. doi:10.3847/1538-4357/aba122

\bibitem[Hunter (2007)]{2007ComputSciEng.9.3} Hunter, J.~D.\ 2007, Computing in Science \& Engineering, vol. 9, no. 3, pp. 90-95

\bibitem[Kylafis(1983)]{1983ApJ...267..137K} Kylafis, N.~D.\ 1983, \apj, 267, 137. doi:10.1086/160851

\bibitem[Lai et al.(2003)]{2003ApJ...598..392L} Lai, S.-P., Girart, J.~M., \& Crutcher, R.~M.\ 2003, \apj, 598, 392. doi:10.1086/378769

\bibitem[Lankhaar \& Vlemmings(2019)]{2019A&A...628A..14L} Lankhaar, B. \& Vlemmings, W.\ 2019, \aap, 628, A14. doi:10.1051/0004-6361/201935064

\bibitem[Lankhaar \& Vlemmings(2020)]{2020A&A...636A..14L} Lankhaar, B. \& Vlemmings, W.\ 2020, \aap, 636, A14. doi:10.1051/0004-6361/202037509

\bibitem[Lankhaar \& Vlemmings(2020)]{2020A&A...638L...7L} Lankhaar, B. \& Vlemmings, W.\ 2020, \aap, 638, L7. doi:10.1051/0004-6361/202038196

\bibitem[Lucy(1971)]{1971ApJ...163...95L} Lucy, L.~B.\ 1971, \apj, 163, 95. doi:10.1086/150748

\bibitem[Mihalas(1978)]{1978stat.book.....M} Mihalas, D.\ 1978, San Francisco: W.H. Freeman, 1978

\bibitem[Morris et al.(1985)]{1985A&A...142..107M} Morris, M., Lucas, R., \& Omont, A.\ 1985, \aap, 142, 107

\bibitem[Mouschovias \& Spitzer(1976)]{1976ApJ...210..326M} Mouschovias, T.~C. \& Spitzer, L.\ 1976, \apj, 210, 326. doi:10.1086/154835

\bibitem[Mouschovias \& Ciolek(1999)]{1999ASIC..540..305M} Mouschovias, T.~C. \& Ciolek, G.~E.\ 1999, The Origin of Stars and Planetary Systems, 540, 305

\bibitem[Panopoulou et al.(2016)]{2016MNRAS.462.1517P} Panopoulou, G.~V., Psaradaki, I., \& Tassis, K.\ 2016, \mnras, 462, 1517. doi:10.1093/mnras/stw1678

\bibitem[Piessens et al.(1983)]{1983qspa.book.....P} Piessens, R., de Doncker-Kapenga, E., \& Ueberhuber, C.~W.\ 1983, Springer Series in Computational Mathematics, Berlin: Springer, 1983

\bibitem[Rasch \& Yu(2003)]{2003JSCIC.25..4} Rasch, J., \& Yu, A.~C.~H.\ 2003, SIAM J. Sci. Comput. Volume 25, Issue 4, pp. 1416-1428 (2003)



\bibitem[Sch{\"o}ier et al.(2005)]{2005A&A...432..369S} Sch{\"o}ier, F.~L., van der Tak, F.~F.~S., van Dishoeck, E.~F., et al.\ 2005, \aap, 432, 369. doi:10.1051/0004-6361:20041729

\bibitem[Stenflo(1994)]{1994ASSL...189} Stenflo, J.~O.\ 1994. Astrophysics and Space Science Library, vol 189. Springer, Dordrecht. https://doi.org/10.1007/978-94-015-8246-96

\bibitem[Skalidis \& Tassis(2021)]{2021A&A...647A.186S} Skalidis, R. \& Tassis, K.\ 2021, \aap, 647, A186. doi:10.1051/0004-6361/202039779

\bibitem[Sobolev(1960)]{1960mes..book.....S} Sobolev, V.~V.\ 1960, Moving Envelopes of Stars, by V. V. Sobolev, Translated by Sergei Gaposchkin, Copyright: 1960, eBook: 2013, Reprint: 2014. Cambridge: Harvard University Press. OCLC: 1013938845. ISBN: 9780674864634, eISBN: 9780674864658.. doi:10.4159/harvard.9780674864658

\bibitem[Townes \& Schawlow(1955)]{1955misp.book.....T} Townes, C.~H. \& Schawlow, A.~L.\ 1955, Microwave Spectroscopy, New York: McGraw-Hill, 1955

\bibitem[Tritsis et al.(2018)]{2018MNRAS.478.2056T} Tritsis, A., Yorke, H., \& Tassis, K.\ 2018, \mnras, 478, 2056. doi:10.1093/mnras/sty1152

\bibitem[Tritsis et al.(2022)]{2022MNRAS.510.4420T} Tritsis, A., Federrath, C., Willacy, K., et al.\ 2022, \mnras, 510, 4420. doi:10.1093/mnras/stab3740

\bibitem[Tritsis et al.(2023)]{2023MNRAS.521.5087T} Tritsis, A., Basu, S., \& Federrath, C.\ 2023, \mnras, 521, 5087. doi:10.1093/mnras/stad829

\bibitem[Troland \& Crutcher(2008)]{2008ApJ...680..457T} Troland, T.~H. \& Crutcher, R.~M.\ 2008, \apj, 680, 457. doi:10.1086/587546

\bibitem[Turk et al.(2011)]{2011ApJS..192....9T} Turk, M.~J., Smith, B.~D., Oishi, J.~S., et al.\ 2011, \apjs, 192, 9. doi:10.1088/0067-0049/192/1/9

\bibitem[van der Tak et al.(2007)]{2007A&A...468..627V} van der Tak, F.~F.~S., Black, J.~H., Sch{\"o}ier, F.~L., et al.\ 2007, \aap, 468, 627. doi:10.1051/0004-6361:20066820

\bibitem[Virtanen et al.(2020)]{2020NatMe..17..261V} Virtanen, P., Gommers, R., Oliphant, T.~E., et al.\ 2020, Nature Methods, 17, 261. doi:10.1038/s41592-019-0686-2

\bibitem[Ward-Thompson et al.(2017)]{2017ApJ...842...66W} Ward-Thompson, D., Pattle, K., Bastien, P., et al.\ 2017, \apj, 842, 66. doi:10.3847/1538-4357/aa70a0

\bibitem[Yang \& Lai(2010)]{2010JTAM..ASRROC..8} Yang, L., \& Lai, S.~P.\ 2010, in JTAM, Vol. 8, ASRROC 2010 Symposium Proceedings: Probing the magnetic field structure in star-forming regions through molecular line
polarization

\end{thebibliography}
%

\appendix
\onecolumn
\section{Statistical equilibrium equations}\label{StEqEq}

Below, we provide the statistical equilibrium equations that describe the interactions between all magnetic sublevels for a system with an arbitrary number of rotational levels $N$

\begin{subequations}
\begin{eqnarray}\label{seeEq1}
\sum\limits_{J=0}^N\sum\limits_{m = 0}^J g_{J, m} n_{J, m} - n_{\text{tot}} = 0
\end{eqnarray}
\begin{eqnarray}\label{seeEq2}
\frac{dn_{J, m}}{dt} = 0 = \sum\limits_{m^\prime = m-1}^{m+1} \frac{g_{J+1, m^\prime}}{g_{J, m}} A_{J+1, m^\prime\rightarrow J, m} n_{J+1, m^\prime} - n_{J, m} \sum\limits_{m^\prime = m-1}^{m+1} f(\Delta g) A_{J, m\rightarrow J-1, m^\prime} + n_{\rm{H_2}}\Bigg(\sum\limits_{J^\prime>J}\sum\limits_{m^=0}^{J^\prime} g_{J^\prime, m^\prime} C_{J^\prime, m^\prime\rightarrow J, m} (n_{J^\prime, m^\prime} - n_{J, m} e^{\Delta E/K_B T)}) \nonumber \\
\ + \sum\limits_{J^\prime<J}\sum\limits_{m^=0}^{J^\prime} g_{J^\prime, m^\prime} C_{J^\prime, m^\prime\rightarrow J, m} (n_{J^\prime, m^\prime} e^{\Delta E/K_B T} - n_{J, m}) + \sum\limits_{m^\prime \neq m} g_{m^\prime} C_{m^\prime\rightarrow m}^{\prime} (n_{J, m^\prime} - n_{J, m})\Bigg) + R_{J+1 \rightarrow J} (n_{J+1, m} - n_{J, m}) - R_{J\rightarrow J-1} (n_{J, m} - n_{J-1, m}) \nonumber \\
\ + \sum \limits_{m^\prime \neq m} \frac{g_{J+1, m^\prime}}{g_{J, m}} U_{J+1\rightarrow J} (n_{J+1, m^\prime} - n_{J, m}) - \sum \limits_{m^\prime \neq m} \frac{g_{J-1, m^\prime}}{g_{J, m}} U_{J\rightarrow J-1} (n_{J, m} - n_{J-1, m^\prime})
\end{eqnarray}
\end{subequations}
where 
\begin{equation}\label{seeEq3}
f(\Delta g) = \begin{cases} 2, & \text{for $\Delta g = 1$} \\
\\
1, & \text{otherwise},
\end{cases}
\end{equation}
where, in turn, $\Delta g = g_{final} - g_{initial}$. In Eqs.~(\ref{seeEq1}--~\ref{seeEq2}) we have implicitly assumed that $m$ and $m^\prime \ge 0$.

\section{Optical depth calculation}\label{SupersedingLVG}

As discussed in \S~\ref{numer}, in order to calculate the optical depth in each cell ($i', j', k'$) of our computational grid, we add the absorption coefficients from all cells ($i, j, k$) that fall within one thermal linewidth (see Eqs~\ref{tauQp} \& \ref{tauQm}). This is a physically-driven choice for physical systems where the LVG is not valid, as photons emitted from one region of such a system will interact with another region that moves with approximately the same velocity. However, when adding the absorption coefficients of cells ($i, j, k$) we use an estimate for their level populations based on the molecular abundance in those cells and the level populations calculated in cell ($i', j', k'$). To demonstrate that this approach does not introduce errors, but it instead drastically improves the accuracy of the calculated level populations compared to the LVG case, we devise the following test for the unpolarized case. For the physical system shown in Fig.~\ref{pyrateinputs} (for which the LVG approximation is not valid) we start by calculating the optical depth in each cell based on the LVG approach; that is the optical depth is only subject to the local physical conditions within each cell. Based on this approach we obtain an ``initial set of level populations'' over the entire computational grid. We then re-iterate over the entire grid and re-calculate the optical depth in each cell ($i', j', k'$) this time considering all grid cells ($i, j, k$) that are one thermal linewidth away. For computing the absorption coefficient in cells ($i, j, k$) we consider their ``initial set of level populations'' from the previous iteration while the level populations and local optical depth in the cell of interest ($i', j', k'$) are allowed to change. We repeat the process, until we achieve convergence simultaneously over the entire grid. In this manner we essentially, consider the coupling between the level populations over the entire grid.

\begin{figure*}
\includegraphics[width=1.\columnwidth, clip]{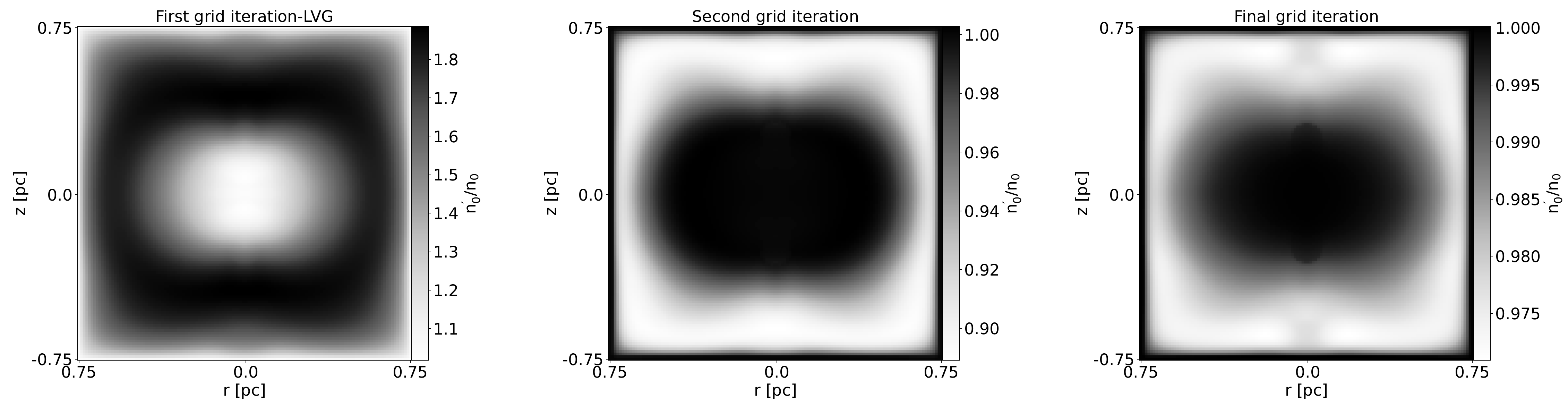}
\caption{Comparison between the population of $J = 0$ calculated using the approach described in \S~\ref{numer} ($n_0$) against the population of $J = 0$ calculated using the methodology described in \S~\ref{SupersedingLVG}. In the latter case, the coupling between the level populations in different regions of the cloud in explicitly taken into account by performing multiple iterations over the entire grid. The method described in \S~\ref{numer} is accurate within 3\% while at the same time the code is $\sim$10 times faster. In both cases, we use as input the physical model shown in Fig.~\ref{pyrateinputs} and use $J\le5$.
\label{popcomp}}
\end{figure*}

In Fig.~\ref{popcomp} we show the compare the population of $J = 0$ using this iterative approach (denoted as $n_0'$) against the population of $J = 0$ computed using the methodology described in \S~\ref{numer}. In the left panel, we show the ratio of the two populations ($n_0'/n_0$) after the first iteration over the entire grid (i.e. LVG-calculated optical depth), in the middle panel, we show the ratio after the second iteration, and in the right panel we show the ratio after the final iteration when the level populations of all rotational levels ($J\le5$) over the entire grid have simultaneously converged (8 iterations with a relative tolerance of $10^{-3}$). As evident from Fig.~\ref{popcomp} using the LVG optical depths leads to more than 40\% errors in the level populations. In contrast, the method described in \S~\ref{numer} is accurate within 3\%, while at the same time the code is $\sim$10 times faster.

Finally, for the a physical system with large enough velocity gradients Eqs~\ref{tauQp} \& \ref{tauQm} clearly reduce to the appropriate limit and the optical depths are calculated under the LVG approximation. That is, for such a system, the level populations in each cell will only depend on the local conditions within the cell and no other cells will be considered in the calculation of the optical depth since they will all be situated more than one thermal linewidth away.

\end{document}